\documentclass[a4paper, twocolumn, 12pt]{article}
\usepackage[utf8]{inputenc}
\usepackage{cite}
\usepackage[centertags]{amsmath}
\usepackage{amsfonts}
\usepackage{textcomp}
\usepackage{amssymb}
\usepackage{amsthm}
\usepackage{caption}
\usepackage{subcaption}
\usepackage[a4paper,left=1.5cm,right=1cm]{geometry}
\usepackage{graphicx}
\usepackage{hyperref}
\usepackage{newlfont}
\usepackage{mathtools, cuted}
\usepackage{lipsum, color}
\usepackage{multicol}
\usepackage{xcolor, soul}
\sethlcolor{green}

\usepackage{xtocinc} 
\usepackage[active]{srcltx} 
\title{\vspace{-2.5cm}\textbf{A sparcely confined water molecules undergoing finite-time thermodynamic processes}}

\author{\small Yigermal Bassie $^a$,Mohammed Mahmud$^b$ and and Mulugeta Bekele$^c$ \\ 
\small $^a$Department of Physics, Wolkite University, Wolkite, Ethiopia.\\
\small $^b$Department of Physics, Wollega University, Wolaga, Ethiopia.\\
  \small $^c$Department of Physics, Addis Ababa University, Addis Ababa, Ethiopia.
}

\begin{document}
\twocolumn[
  \begin{@twocolumnfalse}
\maketitle
\begin{abstract}
\noindent A large number of water molecules are each placed on a lattice far apart so that they are very weakly interacting with each other and in contact with a heat bath at temperature $T$. A strong static electric field, $E_{0}$, is applied to these molecules  along a $z$-axis causing three level split energy values. A weak AC electric field that acts for a finite time $\tau$ applied in the $xy-$plane induces transitions between the three levels. This weak AC field acts as a protocol $\zeta(t)$, that is switched on at $t=0$ and switched off at $t=\tau$. Through this protocol, the system is taken from an initial thermodynamic equilibrium state $F(T,0)$ to the non-equilibrium state $F_{non-equil}(T, \tau)$ recorded right when the AC field is switched off at time $t=\tau$. Once again the AC field is switched on and let it act for the same finite amount of time $\tau$ and its non-equilibrium state $F_{non-equil}(T, \tau)$ recorded right when the AC field is switched off. The same cyclic process is repeated  for a large number of times. The data available for this finite-time non-equilibrium process allowed us to extract equilibrium thermodynamic quantities like free energy, which is what we call Jarznski equality and its relation to the second law of thermodynamics. The work distributions of the three-level system in the optimum condition is obtained. Besides, the average work of the system as a function of $\omega$ and time around the optimum frequency are evaluated, where $\omega$ is the frequency of the AC electric field. \\
\noindent keywords \textit{Non-equilibrium process, Finite-time, Water molecule, Electric dipole, Thermodynamics}
\end{abstract}
\end{@twocolumnfalse}
  ]
\section{Introduction}
Thermodynamics provides macroscopic descriptions of the states of complex systems and their behaviours when they interact or are constrained under various circumstances. It was originally developed to accord with macroscopic systems \cite{callen1985thermodynamics} and is thus based on the idea that a handful of macroscopic variables, such as volume, pressure, and 
temperature, are enough to completely describe a system. Notwithstanding, the beginning of the atomic theory understood the changes in 
the underlying small world to be continually fluctuating with the chaos and randomness of the small-world. In the realm of statistical mechanics the relative fluctuations become insignificant, since it consistently deals with a large number of  particles. Equilibrium statistical mechanics formalism provided a powerful means to explain how the macroscopic properties of many-body systems at thermal equilibrium arise from the microscopic
interactions that occur among their constituent particles. Its fundamental outcome is the Gibbs formula for the canonical 
ensemble \cite{gibbs1902elementary} which mainly employed approach for equilibrium statistical mechanics 
\cite{gibbs1902elementary, feynman1998statistical} to provide a fundamental bridge between microscopic theory and thermodynamic measurements 
for any equilibrium situations. Unlike that of equilibrium statistical mechanics \cite{gibbs1902elementary, feynman1998statistical}, with its well-established foundation, a similar widely-accepted framework for non-equilibrium statistical mechanics \cite{zhong1996effect} 
remains elusive because a handful of parameters used in thermodynamics no longer help to know the whole dynamics of the system; i.e., 
one must study Newton's or Schr$\ddot{o}$dinger's equation for all constituent particles, thus making the problem much more difficult. Indeed, talking about the average behavior of a given ensemble of small systems in statistical mechanics is probabilistic in the sense that as we intend to study deeply the dynamics of the system where micro-state changes always exist fluctuating around the average value as a result of thermal fluctuation, which occurs in a system at equilibrium. These fluctuations are used as the source of noise in many systems.\\

\noindent One of a recent rapidly growing theory deals with the emergence of thermodynamic laws from inherent quantum mechanical theories and its emphasis on the dynamical process out of equilibrium. This new theory is fueled by new, highly controlled quantum experiments
\cite{liphardt2002equilibrium}-
\cite{rousseau1999role}
the discovery of more powerful methods
\cite{schollwock2011density}, and the development 
of new literature tools \cite{miller2017time}, 
\cite{keyl2002fundamentals}.
For instance, the newly developed novel theoretical approaches that help as tools for the development of quantum thermodynamics are non-equilibrium thermodynamics or stochastic thermodynamics \cite{elouard:hal-01170581}.  
Therefore, quantum thermodynamics is a newly emerging research field. The aim of the realm is to extend standard thermodynamics to non-equilibrium statistical mechanics \cite{chou2011non} of small systems. So, researchers look for non-equilibrium conditions to incorporate the whole of quantum effects in this realm.\\

\noindent Research on the non-equilibrium dynamics of quantum systems has deeply produced valuable statements on the thermodynamics of small-scale systems undergoing quantum mechanical progresses \cite{zhong1996effect, schollwock2011density}, \cite{keyl2002fundamentals}-
\cite{crooks2000path}, 
Fundamental examples are produced by the Crooks and Jarzynski relations \cite{jarzynski1997nonequilibrium,jarzynski1997equilibrium, crooks1998nonequilibrium}: taking into account fluctuations in non-equilibrium dynamics, such relations connect equilibrium properties of thermodynamical 
applicability with explicit non-equilibrium features.  In the real world, it is impractical to isolate a particular quantum system, in which we are interested, from its environments. Thus, in order to faithfully represent the real dynamical evolution of physical systems, we must consider the influence of the external environment upon the system's dynamics. Even though the recent advancement in experimental work done on the small systems is effective, we are unable to track either theoretically or experimentally, the dynamical evolution of the full system-plus-environment. This is because the random fluctuations introduced in small systems become valuable and must be incorporated in the explanation of the full system's  dynamics. These random fluctuations in small systems may influence thermodynamic quantities like work and heat. A number 
of authors have proposed definitions of work and derived fluctuation theorems for quantum systems in contact with general thermal environments \cite{miller2017time, talkner2007fluctuation}, \cite{crooks1999entropy, ribeiro2016quantum}. The great insight into the properties of non-equilibrium process could be gained by treating work as a random variable \cite{jarzynski1997nonequilibrium, jarzynski1997equilibrium}. Over time, studies began to look for related conclusions in quantum systems, both for unitary \cite{talkner2007fluctuation} and open \cite{crooks2008jarzynski} quantum dynamics. Recently, in addition to the thermal fluctuations, one more has intrinsically quantum  fluctuations, best to a very richer platform to work with. \\

\noindent Currently, considerable attention has been given to the properties of interacting electric dipole systems. The electrostatic coupling among 
the dipoles makes such systems qualitatively different from their magnetic counterparts. Since interacting spins have been studied during 
last decades, significant progress has been achieved in understanding the underlying physics. In electric dipole systems, the interplay of quantum tunneling, fluctuations and frustration provides with the possibility to realize exotic phases, like quantum electric dipole liquids and glasses \cite{shen2016quantum}, quantum critical phenomena and phase transitions \cite{rowley2014ferroelectric}. Understanding 
the nature of the corresponding phases and their possible relations with magnetic counterparts is of great fundamental and technological 
interest, but is presently still at its infancy.\\

\noindent The rest of the paper is organized as follows.
In section \ref{section2} we explain the experimental arrangement of the system and workout its Hamiltonian and partition function. In section \ref{section3} we describe the general and particular 
procedure of carrying out finite-time cyclic process, formulate the time evolution of the system and define the way to get expectation values of measurable quantities. In section \ref{section4} by taking the finite-time cyclic process of our system we evaluate the probability distribution of work, find its mean value and study the behavior of average work as a function of time. Finally, in section \ref{section5} the current study of the work is summarized and conclusion given. \\
\section{System}\label{section2}
\noindent A water molecule is approximately of 2.75$\mathring{A}$ in diameter. A large number of water molecules can be confined in a nanocavity of beryl hexagonal crystal lattice structure \cite{dressel2018quantum}. One could strongly weaken the dipole-dipole interaction between these molecules by letting them occupy sparcely populated sites, say, every other site emply in a hexagonal lattice structure of 9.2$\mathring{A}$ spacing. Magnitude of the electric dipole of water molecule is found to be $1.85$ Debye, which is $6.17\times 10^{-30}$c-m. One can exert a strong electric field $E_{0}$ in a z-axis of few electrovolts to such a system
\cite{sutmann1998structure}. We attach the system to a heat bath of temperature of $T=300$K. The Hamiltonian of the system containing $N$ such identical confined water molecules  subjected to external electric field $\textbf{E}_{0}$ will be 
\begin{equation}\label{ham21}
 H_{N} = - \sum_{i=1}^{N}\textbf{D}_{i} \cdot \textbf{E}_{0} ,
\end{equation}
where $\textbf{D}_{i}$ is the dipole state of the $i^{th}$ 
water molecule. In the presence of the applied  
external electric field the energy of each confined water molecule will split into three-level energy values. Each energy state of the 
system depends on the orientation of the confined water molecule with respect to the applied electric field.
The confined water molecule takes three energy values when its orientation is aligned parallel ($-D_{i}E_{0}$), anti-parallel ($D_{i}E_{0}$) and perpendicular (0) to the electric field. The Gibbs thermal density operator $ \rho_{th} $, for the system is given by
\begin{equation}\label{dro}
 \rho_{th} = \frac{e^{-\beta H_{N}}}{Z_{N}},
\end{equation}
where the partition function is
\begin{equation}\label{part219}
 Z_{N} = \bigg[1+2\cosh(\beta DE_{0})\bigg]^{N},
\end{equation}
and $\beta = \frac{1}{k_{B}T}$, $k_{B}$ is the Boltzmann constant.
The density operator is successful in explaining the thermal equilibrium states of such system.
 In the next section we explore how the non-equilibrium process of a confined water electric dipole system evolves in the presence of the strong and weak electric fields in a thermal bath.
 \section{Non-equilibrium processes of the system }\label{section3}
 \noindent In this section we first explain the general mode of operatating the cyclic process and point out the particular mode of operation we used. In subsection \ref{322} the time evolution of the system is worked out. Lastly, subsection \ref{3222} deals with evaluating the expectation values of measurable quantities.
\subsection{The cyclic finite-time process of the system}\label{32}
\begin{figure}[h]
 \centering
 \includegraphics[width=8cm,height=4cm]{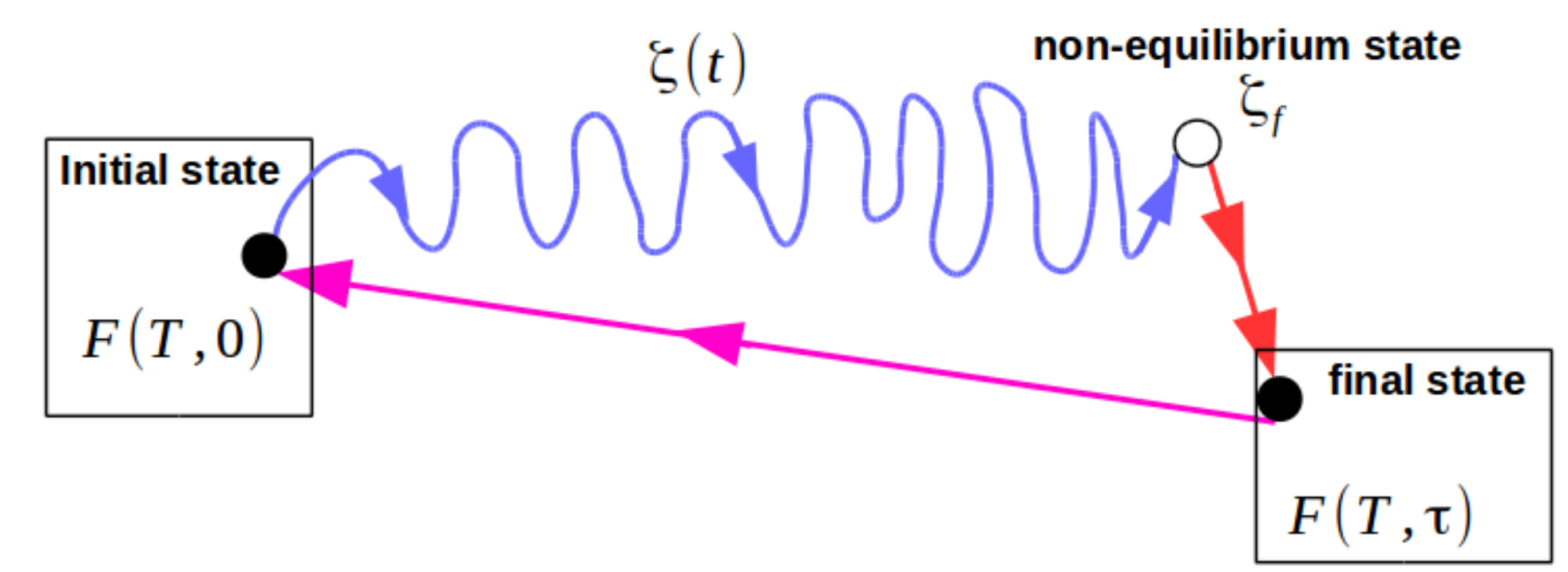}
\caption{ Cyclic finite-time process. Over the protocol $\zeta (t)$,  the system is taken from an original equilibrium state $F (T, 0 )$ to a final non-equilibrium state with parameter $\zeta_{f}$ (blue color line). After the process  is performed, the system will eventually relax from the non-equilibrium state to the final equilibrium state $F(T, \tau )$ (red color line).The reverse path will take the system from its final equilibrium state to its initial equilibrium state in a quasi-static process (pink color line).}
  \label{fig: nonequ3}
 \end{figure}

\noindent In this subsection, we  attach the system to a heat bath and apply a strong electric field, $E_{0}$, along $z$-axis. After the particular confined water electric dipole system stayed enough time to equilibrate with the heat bath, we switch on a weak AC electric field perpendicular to z-direction that lasts for a given amount of time. The weak AC field is the control parameter, $\zeta(t)$, that will act on the system to evolve and make all possible transitions up to time $\tau$. The zig-zag path shown in blue color line in Fig \eqref{fig: nonequ3} depicts the protocol $\zeta(t)$. The dynamics of the system subjected to the AC field will terminate after a span of time $\tau$ at the end of which the final non-equilibrium state, $F_{non-equil}$, measured. After switching off the AC field, but keeping the heat bath and strong electric field in tact, we let the system relax to its final equilibrium state, $F (T, \tau )$. The path from the non-equilibrium state at time $\tau$ (open circle $F_{non-equil}$) to the final equilibrium state (solid circle, $F(T,\tau)$) is shown in red color line in Fig. \eqref{fig: nonequ3}. Once the final equilibrium state is attained, the system will be taken to return back to its initial equilibrium state in a quasi-static process. The reverse path of taking the system from its final equilibrium state to its initial equilibrium state is shown in pink color line in Fig \eqref{fig: nonequ3}.\\
 \\ 
\noindent Once the system has returned to its initial thermodynamic equilibrium state, we initiate the weak AC field in the given direction for the same given span of time $\tau$, measure its final non-equilibrium state at the end, let it relax to its final equilibrium state and, ultimately, return to its initial equilibrium state in a quasi-static process.
This cyclic process will be  performed repeatedly until we get enough data to find the expectation values of any measurable quantities.\\
 
\noindent One important quantity of interest is the amount of work, $W$, performed by the system during this finite-time process. Each measurement of $W$ will, in principle, take different value during each observation. Having a large enough set of measurements will then  enable us to relate it to the change in free energy, $\Delta F$, of the system. In 1997, Jarzynski discovered an equality relation between the work and the change in free energy $\Delta F$  \cite{jarzynski1997nonequilibrium, jarzynski1997equilibrium} which is given by 
\begin{equation}
\langle e^{-\beta W}\rangle =e^{-\beta \Delta F}.
\end{equation}
Note that $\Delta F = F(T,\tau)- F(T,0)$. The above equality is now called Jarzynski equality. A consequence of this equality leads to a relation between $W$ and $\Delta F$ such tat
\begin{equation}
 \langle W \rangle \geq \Delta F.
\end{equation}
The non-equilibrium process disused above is the general mode of operation. However, there are simpler cases when the cyclic process \textit{terminates} after a certain amount of complete cycles. As a result, the value of work will \textit{fluctuate} about zero. And this is the particular mode of operation we considered for our case.
\subsection{The time evolution of the system }\label{322}
\noindent For non-equilibrium process, we can describe the work of the system by requiring detailed knowledge of the dynamics of the system and how it is coupled to the heat bath. In the non-equilibrium process, we assume that the system's coupling to the heat bath is very weak  so that no heat is exchanged with the surrounding. This situation is actually encountered very often in experiments since many systems are only weakly coupled to the bath. It also simplifies considerably the description of the problem because it makes the entire dynamics unitary.\\
\noindent When the weak field, i.e. the protocol of the system is switched on at $t=0$ it has an initial state with Hamiltonian $H_{i} = H(\zeta_{i} )$ in thermal equilibrium with the heat reservoir at a temperature $T$. So the initial state of the system is expressed by using  the Gibbs density matrix in Eq.\eqref{dro}.\\
\noindent Let $E_{n}^{i}$  and $|n \rangle$ denote the eigenvalue and eigenvector of the Hamiltonian $H_{i} = H(\zeta_{i} )$. Then the
 state $|n \rangle$ is obtained with probability
\begin{equation}
P_{n} = \frac{e^{-\beta E_{n}^{i}}}{Z_{N}}.
 \end{equation}
\noindent We  initiate the protocol according to some predefined function $\zeta$ from $\zeta(0) = \zeta_{i}$ to $\zeta(\tau ) = \zeta_{f}$ within the given finite time $\tau$. Due to the very weak coupling of our system with the heat bath, the evolution of the system is unitary. The state of the system at any given time $t$  is given by
\begin{equation}\label{d22}
  |\psi(t) \rangle = U(t) |n \rangle
\end{equation}
where $U(t)$ is the unitary time-evolution operator, which satisfies Schr$\ddot{o}$dinger's equation ($ \hbar= 1$)
\begin{equation}\label{d23}
i\frac{\partial U}{\partial t} = H(t) U,  \textbf{     } U(0) =1.
\end{equation}
At the end of the finite-time process of span $\tau$, we measure the energy of the system. The Hamiltonian at the end of the process, $H_{f} = H(\zeta_{f})$, will be in a given energy level $E^{f}_{m}$ and eigenvector $|m \rangle$. The probability that we measure an energy  $E^{f}_{m}$ is
\begin{equation}\label{d24}
|\langle m|\psi(\tau) \rangle|^{2}= |\langle m|U(\tau)|n \rangle|^{2},
\end{equation}
which can be interpreted as the conditional probability that a system initially in $|n \rangle$ will be found in $|m \rangle$ after a time $\tau$ .\\
\noindent In order to study the out-of-equilibrium properties of this system, we must know the initial thermal state, $\rho_{th}$,  expressed in Eq. \eqref{dro} and the time evolution operator, $U(t)$, which is a solution of the Schr$\ddot{o}$dinger equation  expressed in Eq. \eqref{d23}.\\
\noindent The Hamiltonian $H_{i}$ for each confined water electric dipole induced by an external electric field along the z-axis is described in Eq. \eqref{ham21} and the equivalent matrix 
exponential $e^{\frac{-\hat{H}_{i}}{T}}$ can be evaluated by exponentiating the eigenvalues
\begin{equation}
  e^{\frac{-\hat{H}_{i}}{T}} =
  \begin{pmatrix}
 e^{\frac{DE_{0}}{T}} & 0 & 0 \\
 0 & 1 & 0\\
 0 & 0 & e^{\frac{-DE_{0}}{T}}
 \end{pmatrix}.
 \end{equation}
The trace of this matrix is the partition function for the single water molecule
\begin{equation}
  Z = 1+2\cosh(\frac{DE_{0}}{T}).
 \end{equation}
\noindent The thermal density matrix can be written in the form of
\begin{equation}\label{dens1}
 \rho_{th} =
  \begin{pmatrix}
\frac{1-f}{2} & 0 & 0 \\
0 & \frac{f}{1-2\sinh(\frac{DE_{0}}{T})} & 0\\
0 & 0 & \frac{1+f}{2}-\frac{f}{1-2\sinh(\frac{DE_{0}}{T})}
\end{pmatrix},
 \end{equation}
where $f=\frac{1-2\sinh(\frac{DE_{0}}{T})}{1+2\cosh(\frac{DE_{0}}{T})}$.\\
\noindent Next, we obtain the  analytical solution of the Schr$\ddot{o}$dinger equation for time-dependent Hamiltonian. Fortunately, in our case one may obtain an approximate solution valid when $E_{I} \ll DE_{0}$ and take the direction of the weak electric field to be rotating in the $xy$-plane. The work protocol is enforced by applying a very small field of amplitude $E_{I} $ rotating in the $xy$-plane with  frequency $\omega$. Therefore, the work parameter $\zeta$ is defined by the field $\textbf{E}_{I}  = \textbf{E}_{I}( \cos\omega t, \sin\omega t, 0)$. The Hamiltonian describing the response of a single confined water molecule as an electric dipole system to the two fields is given by 
\begin{equation}
\hat{H} = -DE_{0}\boldsymbol{\hat{\sigma}_{z}} -\textbf{E}_{I} (\boldsymbol{\hat{\sigma}_{x}}\cos\omega t + \boldsymbol{\hat{\sigma}_{y}}\sin\omega t),
 \end{equation}
where, $\boldsymbol{\hat{\sigma}_{x}}$, $\boldsymbol{\hat{\sigma}_{y}}$ and $\boldsymbol{\hat{\sigma}_{z}}$ are the Pauli matrices for spin-1 particles.
Next, we compute approximate solution of the time evolution operator $U(t)$ defined in \eqref{d23} and let us first define a new operator such that
\begin{equation}\label{operator13}
U(t)=\tilde{U}(t)e^{i\omega t\boldsymbol{\hat{\sigma}_{z}}}.
\end{equation}
Substituting Eq.\eqref{operator13}) into Eq.\eqref{d23}, one finds that $\tilde{U}$ must obey the modified Schr$\ddot{o}$dinger equation \begin{equation}\label{modshero13}
i\frac{ \partial \tilde{U}(t) }{\partial t} = (H+\omega \boldsymbol{\hat{\sigma}_{z}}) \tilde{U} =\tilde{H}\tilde{U}(t),
 \end{equation}
 where
\begin{equation}\label{heol137}
\tilde{H}= (\omega-DE_{0})\boldsymbol{\hat{\sigma}_{z}} -\textbf{E}_{I} (\boldsymbol{\hat{\sigma}_{x}}\cos\omega t + \boldsymbol{\hat{\sigma}_{y}}\sin\omega t).
\end{equation}
Therefore, from the expression (\ref{heol137}), we have two distinct types of terms, one time-independent and the other oscillating with frequency $\omega$. The system consists of a single spin placed with a static electric field in the $z$ direction, together with a weak oscillating field $\textbf{E}_{I} = A \sin \omega t$ of frequency $\omega$ in the $xy$ plane, which plays the role of the work control parameter $\zeta$. Now the expression (\ref{heol137}) can be rewritten as
\begin{equation}\label{yig23}
\tilde{H} = (\Omega \boldsymbol{\hat{\sigma}_{z}}-b\sigma_{y}) -b(\boldsymbol{\hat{\sigma}_{x}}\sin 2\omega t -\boldsymbol{\hat{\sigma}_{y}}\cos2\omega t),
\end{equation}
where $\Omega =\omega-DE_{0}$. After neglecting any time-dependent terms from the expression in Eq. (\ref{yig23}), we are left only with the much simpler Hamiltonian, which is 
\begin{equation}\label{rewrlneg13y}
\tilde{H} = \Omega \boldsymbol{\hat{\sigma}_{z}}-b\boldsymbol{\hat{\sigma}_{y}}. 
\end{equation}
Express (\ref{rewrlneg13y}) as Thus, we can  apply the trick to our issue by writing Eq. (\ref{rewrlneg13y}) as
\begin{equation}\label{rewrlneg13}
\tilde{H} = \Omega_{r}(\boldsymbol{\hat{\sigma}_{z}}\cos\theta - \boldsymbol{\hat{\sigma}_{y}}\sin\theta) 
\end{equation}
where 
\begin{equation}\label{reyig}
\Omega_{r} = \sqrt{\Omega^{2}+b^{2}}     \text{     and     } \tan\theta = \frac{b}{\Omega}.
\end{equation}
After writing an explicit formula for the full time-evolution operator defined in Eq. \eqref{operator13}, we must compute the exponential matrix $e^{-i\tilde{H}t}$. Rearranging  the terms in the exponential, the unitary time-evolution operator can be expressed as
\begin{equation}\label{res1}
U(t) =
 \begin{pmatrix}
 u(t) & y^{\ast}(t)& w^{\ast}(t) \\
 -v^{\ast}(t) &x(t)& v(t)\\
 w(t) & - y(t)& u^{\ast}(t)
 \end{pmatrix},
\end{equation}
 where the amplitude probabilities are:
\begin{equation}\label{operateru}
u(t) =  e^{i\omega t \boldsymbol{\hat{\sigma}_{z}}}e^{-i\Omega_{r} t\cos\theta}\frac{1}{2}\{1+\cos(\Omega_{r} t \sin\theta)\},
\end{equation}
\begin{equation}\label{reshq1}
 v(t) = e^{i\omega t \boldsymbol{\hat{\sigma}_{z}}}\frac{\sqrt{2}}{2}\sin(\Omega_{r} t \sin\theta),
\end{equation}
\begin{equation}\label{reshq2}
w(t) = e^{i\omega t \boldsymbol{\hat{\sigma}_{z}}} e^{i\Omega_{r} t\cos\theta}\frac{1}{2}\{1-\cos(\Omega_{r} t \sin\theta)\},
\end{equation}
\begin{equation} \label{reshq3}
y(t) = e^{i\omega t \boldsymbol{\hat{\sigma}_{z}}} e^{-i\Omega_{r} t\cos\theta}\frac{\sqrt{2}}{2}\sin(\Omega_{r} t \sin\theta),
\end{equation}
\begin{equation}\label{operateru1}
x(t)=  e^{i\omega t \boldsymbol{\hat{\sigma}_{z}}}  \cos(\Omega_{r} t \sin\theta).
\end{equation}
To get a better physical interpretation of this result, consider the situation where the system initially starts in the eigenstate $ | 1,1 \rangle$ of $\boldsymbol{\hat{\sigma}_{z}}$ . Then $y^{\ast}(t)$ and $w^{\ast}(t)$, being the off-diagonal elements of $U (t)$, describe the probability amplitude for a transition from  $| 1,1 \rangle$ $\rightarrow$ $ | 1,0 \rangle$ or $ | 1,-1 \rangle$ (the amplitude for the reverse transition processs  are $v^{\ast}(t)$ and $w(t)$ ). Moreover, the unitarity condition $U^{\dagger}(t) U(t) = \mathbb{1}$ also implies that $ |u(t)|^{2} + |y(t)|^{2} +|w(t)|^{2}= 2|v(t)|^{2} + |x(t)|^{2} = 2|y(t)|^{2} + |x(t)|^{2} =|u(t)|^{2} + |v(t)|^{2} +|w(t)|^{2} =1$. Then, the expressions $|u(t)|^{2}$ and $|x(t)|^{2}$ are the probabilities for no transitions to occur.\\
\noindent The transition probability reaches a maximum precisely at an optimum condition ($\Omega = 0$), as we intuitively expect. In fact, at optimum condition, we obtain the  transition probabilities such as $|v(t)|^{2}$, $|w(t)|^{2}$ and $|y(t)|^{2}$ achieve maximum value. Hence, 
when the weak oscillating field is applied along $xy$-plane\textbf{ transitions do occur}.
\subsection{ The expectation value of measurable quantities }\label{3222}
To illustrate the physics behind Eq \eqref{res1}, let us examine the time evolution of the mean polarization components $\langle \sigma_{x} \rangle$,  $\langle \sigma_{y} \rangle$, and $\langle \sigma_{z} \rangle$. The general formula for the time evolution of the mean of any operator, $A$, is given by
\begin{equation}\label{dynamics}
\langle A \rangle = tr \{ U^{\dagger}(t) A U(t) \rho_{th}\}.
\end{equation}
Now, we obtain the time evolution of the mean polarization components in the three different directions,  $\langle \sigma_{i} \rangle$, where $i=x, y, z$.  By using Eq. \eqref{dynamics}, we can obtain the mean polarization components in the $x, y,$ and $z$ directions to be
\begin{equation}\label{sigmmay1_sample}
\langle \sigma_{x} \rangle = -\frac{2\sin(\Omega_{r}t\sin(\theta))\cos(\Omega_{r}t\cos(\theta)) \sinh(\frac{DE_{0}}{T})} {1+2\cosh(\frac{DE_{0}}{T})},
\end{equation}
\begin{equation}
\begin{split}
\langle \sigma_{y} \rangle = - \frac{2\sin(\Omega_{r}t\cos(\theta))\sin(\Omega_{r}t\sin(\theta))\sinh(\frac{DE_{0}}{T})} {1+2\cosh(\frac{DE_{0}}{T})},
 \end{split}
\end{equation}
 and
\begin{equation}\label{sigmaza}
\langle \sigma_{z} \rangle = \frac{2\cos(\Omega_{r}t\sin(\theta))\sinh(\frac{DE_{0}}{T})}{1+2\cosh(\frac{DE_{0}}{T})}.
\end{equation}
\noindent The full expressions using Eqs. \eqref{operateru}-\eqref{operateru1} are somewhat bulky. Instead, let us look at the \emph{optimum} case, where the dynamics of the three-level system of spins have a simplified form of the three-level which turn out to be 
 \begin{equation}\label{mes12}
\langle \sigma_{z} \rangle_{opt} = -\frac{2\cos(b^2t)\sinh(\frac{DE_{0}}{T})}{1+2\cosh(\frac{DE_{0}}{T})},
\end{equation}
\begin{equation}
\langle \sigma_{x}\rangle_{opt}  =-\frac{2\sin(b^2t)\sinh(\frac{DE_{0}}{T})}{1+2\cosh(\frac{DE_{0}}{T})},
\end{equation}
and
\begin{equation}\label{mes123}
 \langle \sigma_{y} \rangle_{opt} =0.
\end{equation}
\noindent The expectation values of measurable quantities at the optimum condition expressed in Eqs.\eqref{mes12}-\eqref{mes123}, are the possible expected values of the result of the measurable quantities. Note that this optimum condition for the mean polarization along $y$ direction, $\langle \sigma_{y}\rangle$, expressed in Eq. \eqref{mes123}  has zero probability of occurrence.
\section{Exploring work distribution properties of the system } \label{section4}
\noindent This section deals with the result and discussion. Subsection \ref{41} presents the work distribution of our system. In subsection \ref{42} average work of the three-level system will be evaluated with some results explained. In subsection \ref{44} the average work of our system as a function of time is derived along with some results explained.
\subsection{ Work distribution of the system}\label{41}
We begin our study of the considered system by defining a two point energy values which is the difference of the two point energy measurement of the system given by
 \begin{equation}\label{rework}
  W = E_{m}^{f} -E_{n}^{i}.
 \end{equation}
Due to the very weak interaction of the system with the heat bath heat exchange with the bath will be neglected. But, any change in the energy value of the system must be related to the work performed by the external agent. We denote the energy recorded in the first measurement to be $E_{n}^{i}$ while the energy value recorded at the end of each finite-time $\tau$  to be $E_{m}^{f}$. These energy  $E_{m}^{f}$ measured at the end of each realization fluctuate subjected to the quantum evolution of the system. On the other hand, the initial measurement of energy  $E_{n}^{i}$ is random due to thermal fluctuation. As a result, $W$ can be treated as a random variable, encompassing both thermal and quantum fluctuations during each realization of the measurement. From Eq. \eqref{rework} we recognize that work is a quantity which requires two measurements to be accessed. This reflects to the fact that work is not the system property, but rather the result of a process performed on the system.\\
Since the collection of the particles are assumed to be weakly interacting with each other compared to each particle's interaction with the external field, we can take the state of each particle to be determined by  the interaction Hamiltonian with the external fields. Hence the study of our system can boil down to simply observing the state of a single representative spin-one particle. And that is what we will do in the following work.\\
Each spin-one particle has three energy states and one can figure out all possible values of work ($W$). At time $t = 0$ the Hamiltonian is $H_{i}= -DE_{0}\sigma_{z}$ so the initial energy eigenvalues can take one of the following three: 
$E^{i}_{\pm 1} = \mp DE_{0}$ and $E^{i}_{0} = 0$. At some other arbitrary time the Hamiltonian is $H(t)= -DE_{0}\sigma_{z}-E_{I}(t) \sigma_{x}$ so the final instantaneous eigenvalue takes any one of the following: $E^{f}_{\pm 1} = \mp \sqrt{(DE_{0})^{2} + E_{I}^{2}(\tau)}$ and $E^{f}_{0} = 0$. From here, there are nine possible values of $W$:
 \begin{equation}
  W=E^{f}_{m}-E^{i}_{m},
 \end{equation}
where $m=\pm 1$ and $0$.\\
In order to simplify the discussion, let us suppose that we choose the protocol such that $E_{I} = b\sin(\omega t)$
always changes by a full period. That is, we assume that the final protocol time $\tau$ is an \emph{integral} multiple of the period of the weak field, i.e.
 \begin{equation}
  \tau = \frac{2 \pi l}{\omega},  \textbf{   } l = 1, 2, 3, .....
 \end{equation}
This is physically quite reasonable. After all, $\omega$ is supposed to be of very high frequency and therefore we imagine always measuring the work after a certain amount of complete cycles. As a result, the value of work should be fluctuating about zero. \textit{For this particular mode of operation, we note that the reverse path is not required as the system has already returned to the same condition excepting the finite-time process}. \\
With this choice $H_{f} = H_{i}$ we study the quantum thermodynamic properties of a three-level system, so they have energy spectrum $E_{m=\pm 1} = \mp DE_{0}$ and $E_{m=0} = 0$. 
Then, we obtain the following possible distributions of work for the single particle in the system:
 \begin{equation}\label{woq}
 \begin{split}
  W =E_{-1} - E_{1} = 2DE_{0} \\      
  W= E_{1} - E_{-1} =-2DE_{0}     \\
  W= E_{0} - E_{1}= DE_{0}          \\
  W= E_{1} - E_{0} = -DE_{0}       \\
  W= E_{-1} - E_{0} = DE_{0}         \\
  W= E_{0} - E_{-1}= -DE_{0}           \\
  W= E_{0} - E_{0} = 0                   \\
  W= E_{1} - E_{1} = 0                 \\
  W= E_{-1} - E_{-1} =0 
  \end{split}   
  \end{equation}
In the first case, the spin  initially was found in state $m=1$ at $t=0$ and then found in state $m=-1$ at $t=\tau$.
The second case, correspond to the reverse process of the first case. In the third case, the spin state changes from state $m=1$ to the state $m=0$ and the fourth correspond to the reverse process of the third state. The fifth and sixth cases, correspond to the transition of state $m=0 (-1)$ to the state  $m=-1 (0)$ respectively. The last three expressions correspond to no state transition at all. \\
The probability distribution of work, $P(W)$, can be evaluated by using the definition
\begin{equation}\label{s30}
 P(W) = \sum_{n, m}|\langle m|U(\tau)|n\rangle|^{2}P_{n}\delta[W-(E^{f}_{m}-E^{i}_{n})],
\end{equation}
where $\delta(x)$ is the Dirac's delta function and $x=W-(E^{f}_{m}-E^{i}_{n})$. This expression is  explained in words as the sum over all allowed events, weighted by their probabilities, and catalogue the terms according to the values of $E^{f}_{m}-E^{i}_{n}$.\\
For instance, the case $W = 2DE_{0}$  means a transition from quantum state $m=1$ to quantum state $m=-1$. From Eq. \eqref{dens1} we have the initial probability $P_{1}= (1-f )/2$, whereas the the  transition probability  is $|m=-1|U (\tau)|m=1|^{2} =|w(\tau)|^{2} $. Therefore, we obtain all the possible transition probabilities to be given as follows:
 \begin{multline}\label{brobdistr}
P(2DE_{0})=[ \frac{1-f}{2}] |w(\tau)|^{2}          \\
P(-2DE_{0})= [\frac{1+f}{2} -\frac{f}{1-2\sinh(DE_{0}/T)}] |w(\tau)|^{2}   \\
P(DE_{0}) =  [ \frac{1-f}{2}] |y(\tau)|^{2}      \\
P(-DE_{0})  =   [\frac{f}{1-2\sinh(DE_{0}/T)}] |v(\tau)|^{2}   \\
P( DE_{0})  =  [\frac{f}{1-2\sinh(DE_{0}/T)}] |v(\tau)|^{2}   \\
P(-DE_{0})=  [\frac{1+f}{2} -\frac{f}{1-2\sinh(DE_{0}/T)}] |y(\tau)|^{2}  \\
P(0)=  1-\{(1-\frac{f}{1-2\sinh(DE_{0}/T)})|w(\tau)|^{2} \\
 + [\frac{1-f}{2} +\frac{f}{1-2\sinh(DE_{0}/T)}] |y(\tau)|^{2} \}        \\
P(0)= 1-\{ (1-\frac{f}{1-2\sinh(DE_{0}/T)})|v(\tau)|^{2} \}                 \\
P(0)=  1-\{(1-\frac{f}{1-2\sinh(DE_{0}/T)})|w(\tau)|^{2} \\
  +    (\frac{1+f}{2}) |v(\tau)|^{2} \}           
 \end{multline}
\subsection{Average work of the three-level system} \label{42}
Once we know the probability distribution of work, $P(W)$, we can get the average work of each representative spin-one particle from the definition:
\begin{equation}\label{proba4}
 \langle W\rangle = \sum_{W} W P(W).
\end{equation}
Using Eq.(\ref{brobdistr}) and Eq.(\ref{proba4}) we obtain the simplified average  work of the three-level system as
\begin{multline}
\langle W\rangle = \frac{-4DE_{0}\bigg(\sin(\Omega_{r}t\sin(\theta))\bigg)^{4}\cosh(\frac{DE_{0}}{T})}{ 8+16\cosh(\frac{DE_{0}}{T})}\\+\frac{2DE_{0}\bigg(\cos(\Omega_{r}t\sin(\theta)-1)\bigg)^{2}\sin(\frac{DE_{0}}{T})}
 { 8+16\cosh(\frac{DE_{0}}{T})}\\
  +\frac{DE_{0}\bigg(1+cos(\Omega_{r}t\sin(\theta))\bigg)^{4}}
  {8+16\cosh(\frac{DE_{0}}{T})}.
 \end{multline}
 \begin{figure*}[ht]
\begin{subfigure}{.5\textwidth}
  \centering
  \includegraphics[width=.8\linewidth]{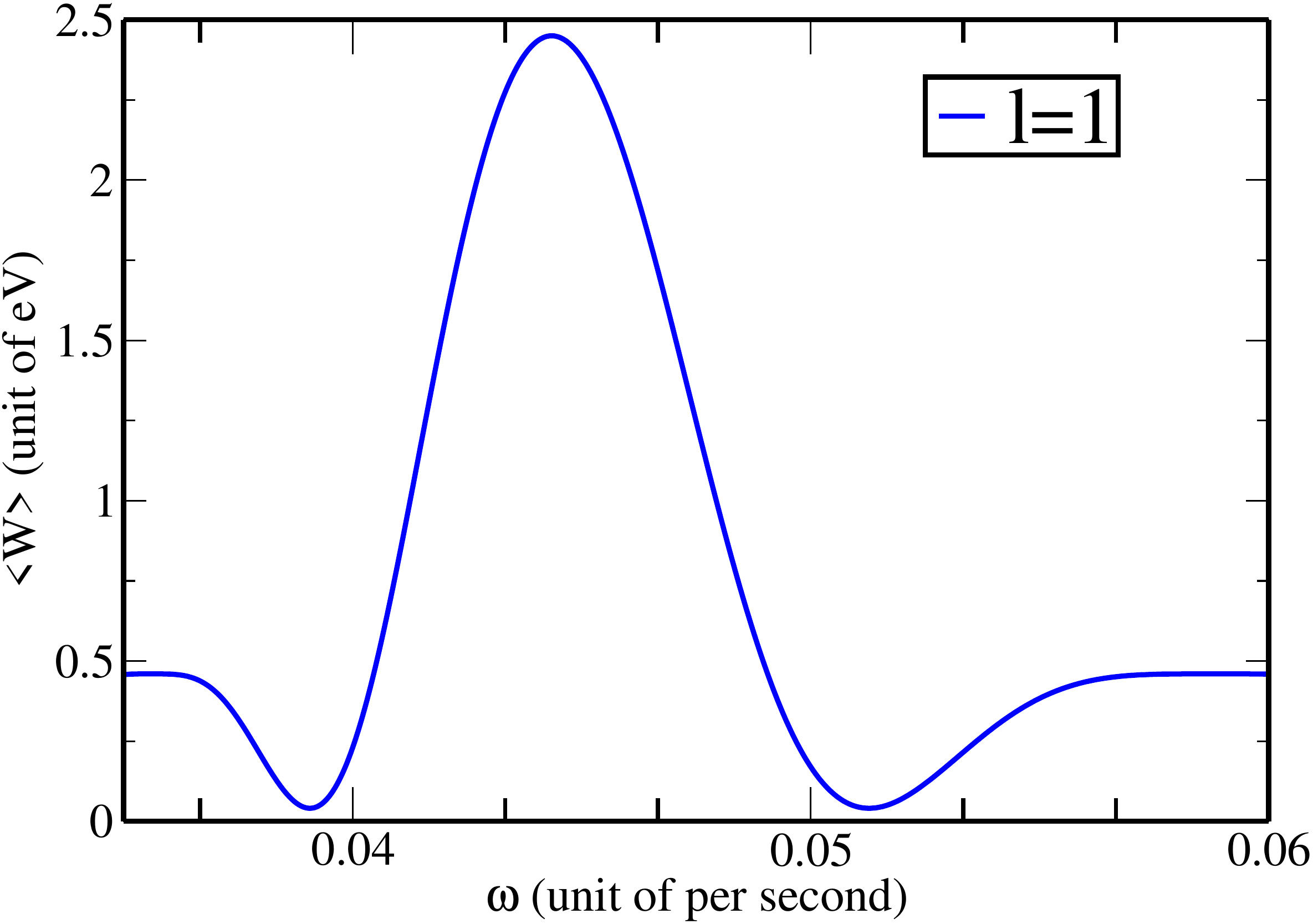}  
  \label{fig:sub-first}
\end{subfigure}
\begin{subfigure}{.5\textwidth}
  \centering
  \includegraphics[width=.8\linewidth]{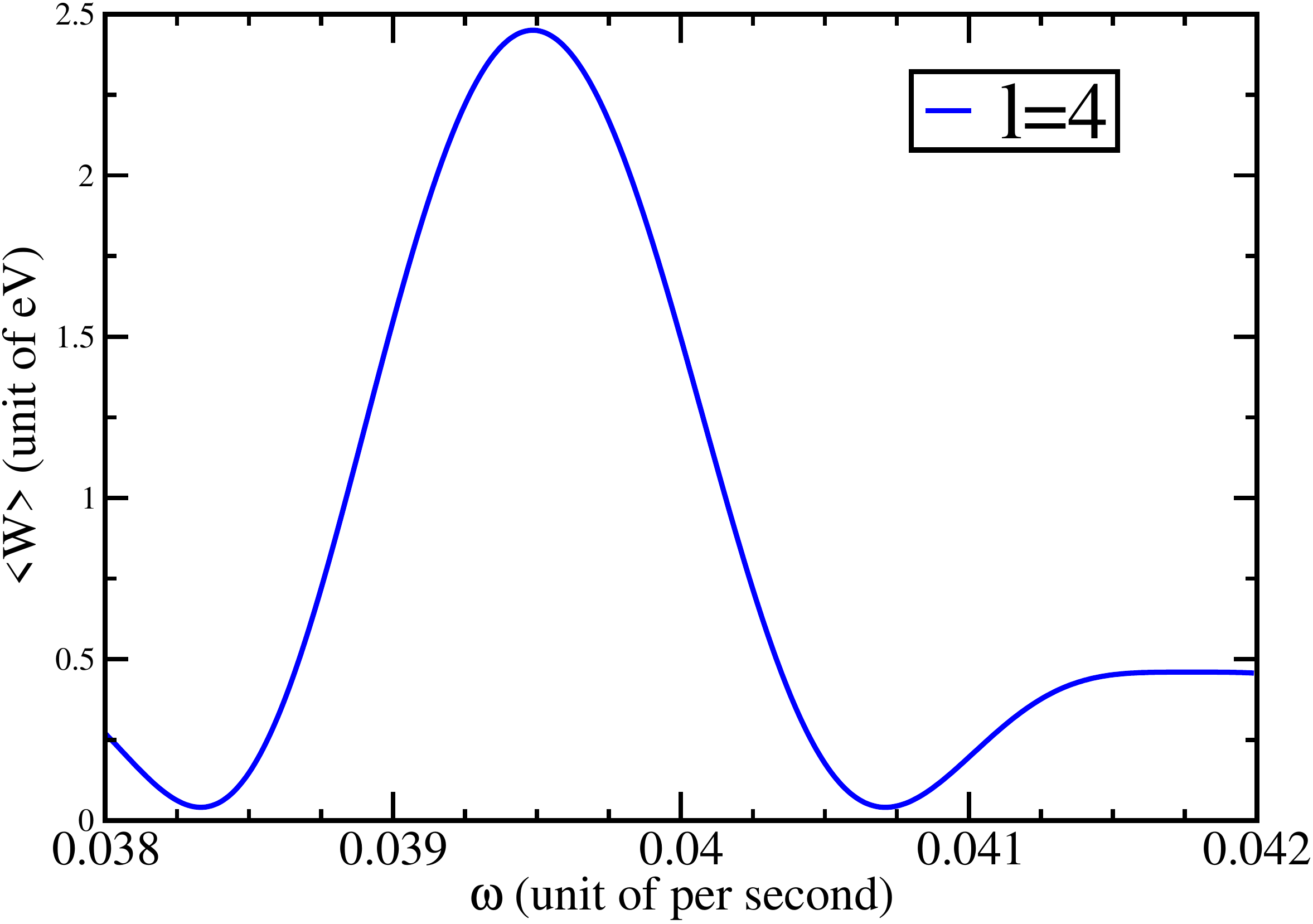}  
  \label{fig:sub-second}
\end{subfigure}
\newline
\begin{subfigure}{.5\textwidth}
  \centering
  \includegraphics[width=.8\linewidth]{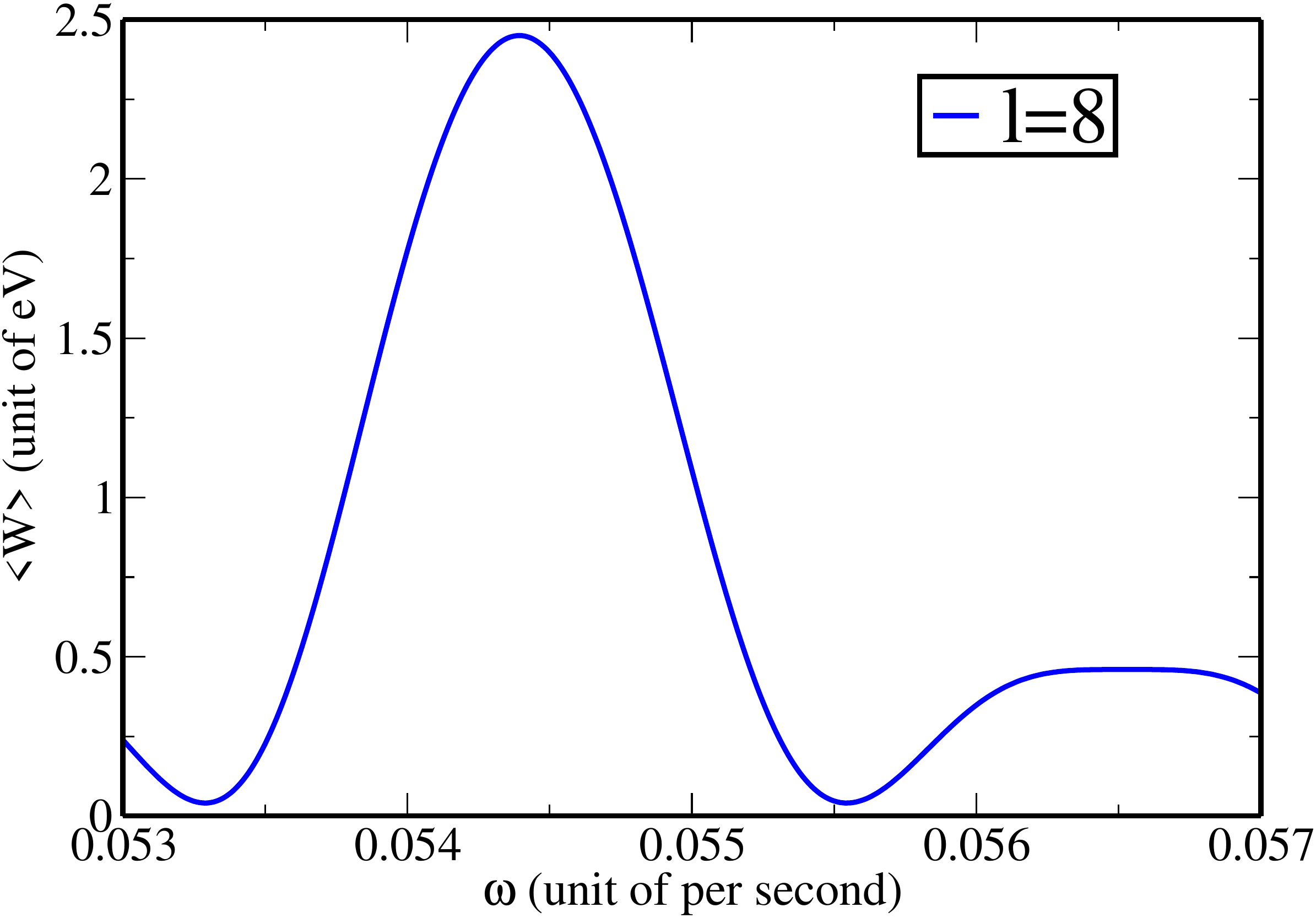}  
  \label{fig:sub-third}
\end{subfigure}
\begin{subfigure}{.5\textwidth}
  \centering
  \includegraphics[width=.8\linewidth]{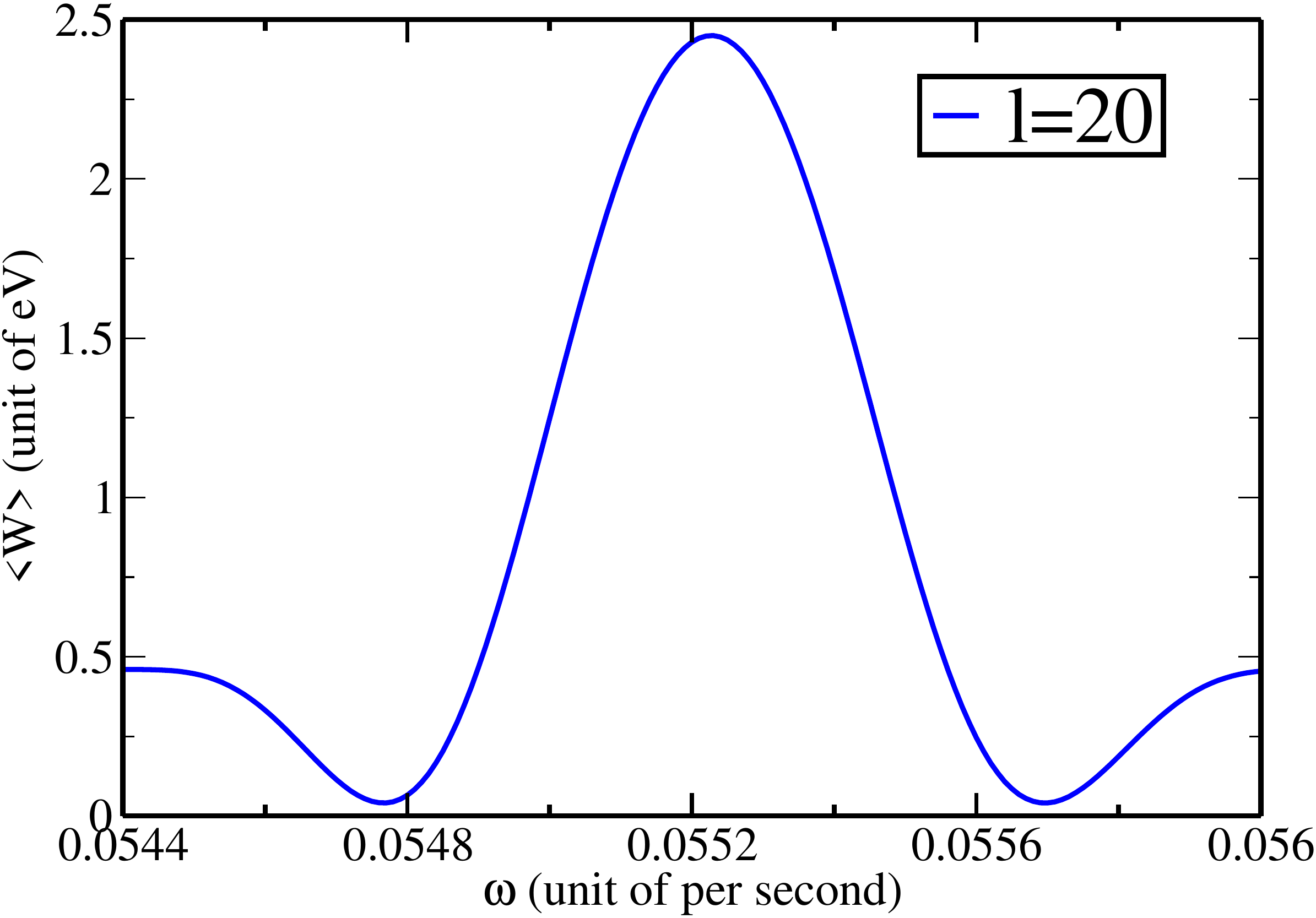}  
  \label{fig:sub-fourth}
\end{subfigure}
\caption{The plot of average work as function of $\omega$ at the values of $DE_{0}=3 eV$, $b=0.01 DE_{0}$, $T=300K$ and different values of $l$.}
\label{fig91}
\end{figure*}
\noindent Figure \ref{fig91} are plots of average work vs frequency of the three-level system for different values of $b/DE_{0}$ and $l$ in a narrow window of frequency. As we see the  dependence of $\langle W\rangle$ on $\omega$ is quite complicated - highly non-linear- and depends sensitively on the duration $l$ of the protocol and the static electric field. From the plots we see the dependence of average work on the strength of strong static electric field and the parameter, $l$. In the  presence of strong static electric field the system performs a maximum average work around the optimum condition. In this optimum condition, the maximum average work of the system depends on the strength of the static electric field. In this condition, the performed average work increases as the static electric field increases. But in a certain value of the parameter, $l$, we have a similar behavior of the average work. As we increase the value of the parameter, $l$, of the system we see a sharp dependence of the average work with frequency, $\omega$, being maximum at the optimum condition. In general, the work performed increases close to the optimum condition, but for certain value of $l$ it may be very small exactly at optimum condition.
\subsection{ Average work of the three-level system as a function of time} \label{44}
The average work as a function of time of the three-level systemcan be given by the formula:
\begin{equation}\label{WasT}
  \langle W \rangle_{t} = \langle H_{f} \rangle_{\tau} -\langle H_{i}\rangle_{0}.
 \end{equation}
 When calculating the expected values of quantities related to the system energy at the time $t = 0$, 
 we can use the initial Hamiltonian $ H_ {0} = -DE_{0} \sigma_ {z} $ instead of the whole time dependent 
 Hamiltonian $ H (t) $ in Eq. \eqref{yig23}.
 The Hamiltonian at time $t$ is given by
 \begin{equation}
  H(t) = -DE_{0} \sigma_{z} -2b \sin \omega t \sigma_{x}.
 \end{equation}
 The average Hamiltonian of the system at any given time can be found from Eq. \eqref{dynamics} where  $A=H(t)$
 \begin{equation}
 \langle H_{f} \rangle_{\tau}= -DE_{0}\langle \sigma_{z}\rangle -2b \sin \omega t \langle \sigma_{x}\rangle
 \end{equation}
 Using Eqs. \eqref{sigmaza} and \eqref{sigmmay1_sample} substitute in to Eq.\eqref{WasT} and rearranging it we find the average work of the three-level system as a function of time to be
\begin{multline}\label{av_w_as_time}
\langle W \rangle_{t} =
 \frac{4[b\sin \left( \omega\,t \right) \sin \left( \Omega_{r}\,t\sin
  \left( \theta \right)  \right) \cos \left( \Omega_{r}\,t\cos \left( 
\theta \right)  \right)}
 { 1+2\,\cosh \left( {\frac {{ DE_{0}}}{T}}\right)}\\-\frac{\frac{ DE_{0}}{2} \left( \cos \left( \Omega_{r}\,
 t\sin \left( \theta \right)  \right) -1 \right) ]\sinh \left( {\frac {
 { DE_{0}}}{T}} \right)} 
 { 1+2\,\cosh \left( {\frac {{ DE_{0}}}{T}}\right)} 
\end{multline}
 \begin{figure}[h]
  \centering
  \includegraphics[width=8cm,height=8cm]{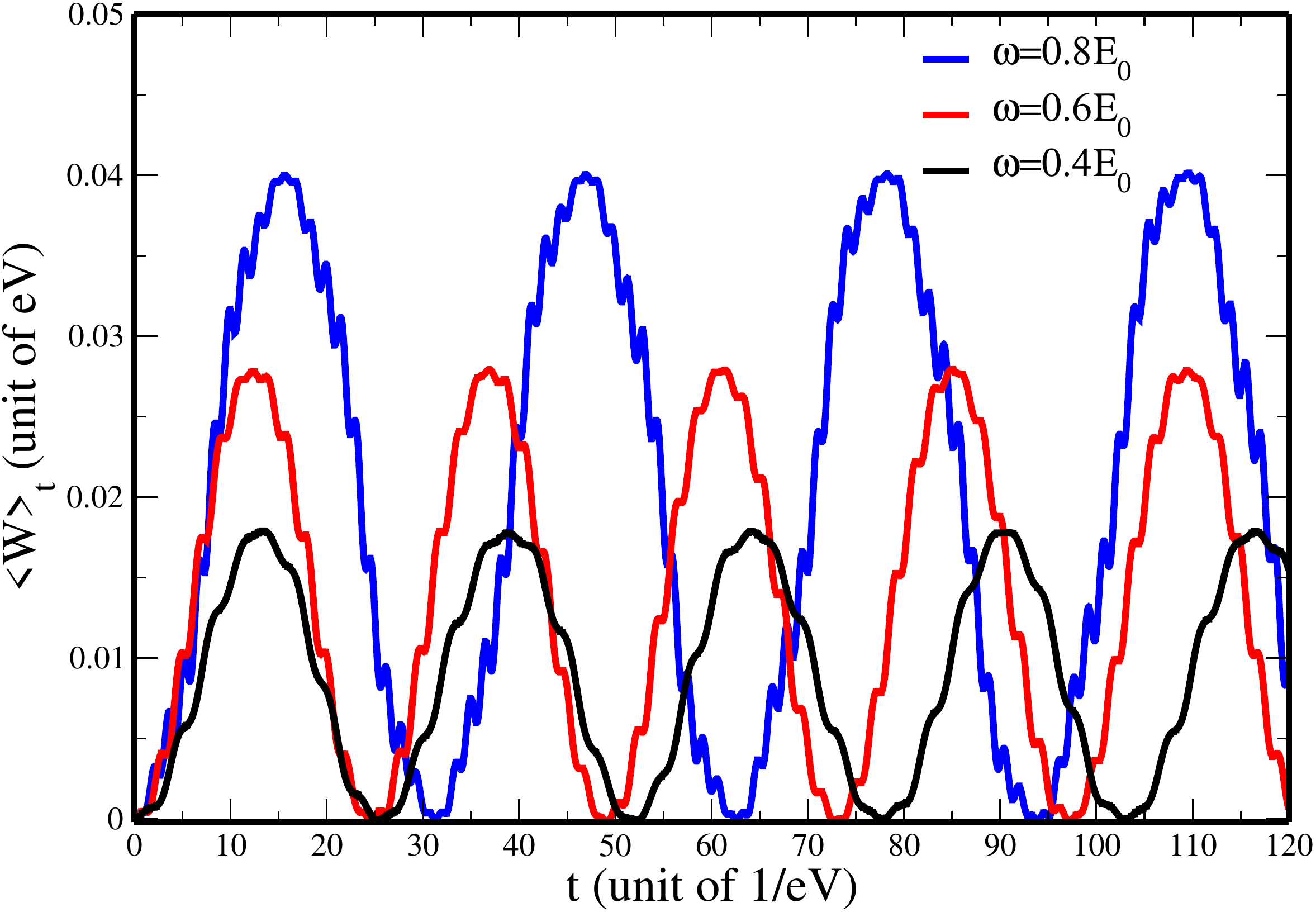}
 \caption{The average work as a function of time for the values of $DE_{0}= 3eV, 2.5eV$ and $2eV$ and the values of $b =0.1DE_{0}$.}
 \label{fig:av work as of time1}
 \end{figure}
\noindent  Figure \eqref{fig:av work as of time1} shows three  plots where $DE_{0}$ takes values of $3eV$, $2.5eV$ and $2eV$ while $\omega$ correspondingly takes $0.8 DE_{0}$, $0.6 DE_{0}$, $0.4 DE_{0}$, respectively for $b=0.1DE_{0}$. The temperature of the heat bath is taken to be $300K$. All the three plots show oscillatory behavior where the values of the average work as a function of time takes between
 zero and maximum value. 
The higher $DE_{0}$ is the higher is the corresponding maximum value. One also observes two time scales: fast and slow oscillations, where the fast oscillation should be related with $\omega$ while the slow oscillation should
be related to the value of $b$. The slow time scale enslaves the dynamics and, as such, is responsible for governing the oscillatory nature of average work as a function of time. 
 \begin{figure}[h]
  \includegraphics[width=8cm,height=8cm]{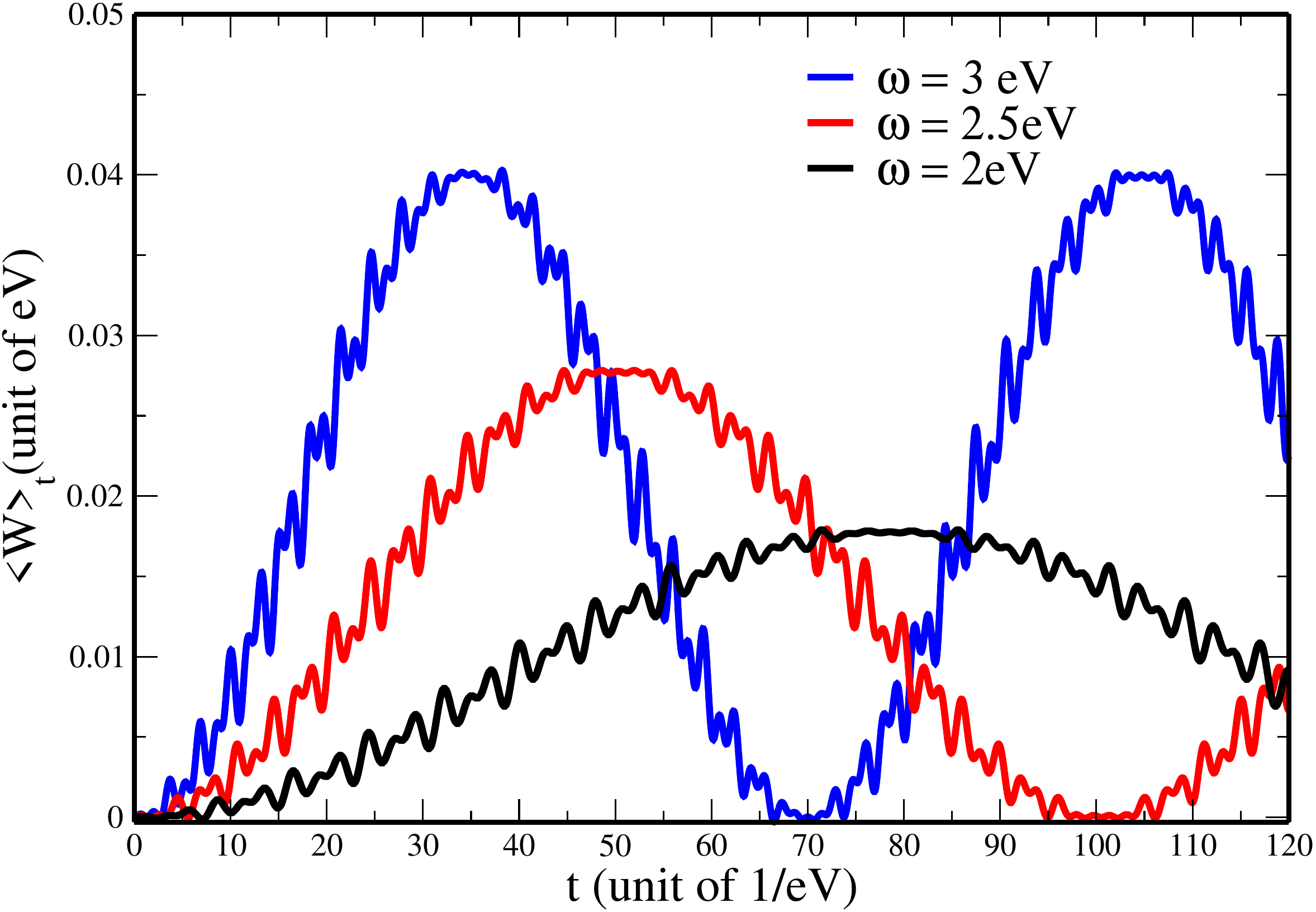}
\caption{The average work as a function of time when $\omega = DE_{0}$ for the values of $DE_{0}= 3eV, 2.5eV$ and $2eV$ and the values of $b =0.1DE_{0}$.}
  \label{fig:jhk1q1}
 \end{figure}
Figure \eqref{fig:jhk1q1} shows another three plots where $DE_{0}$ takes values of $3eV$, $2.5eV$ and $2eV$ while $\omega$ takes  corresponding fixed values of $DE_{0}$ for $b=0.1DE_{0}$. Temperature, $T$, is again taken to be $300K$. The plots here  show clearly the fast and slow oscillations. The fast time scale corresponds to $\omega = DE_{0}$ while the slow time scale corresponds to $b = 0.1DE_{0}$. In addition, the slow oscillation determines the oscillatory behavior of average work as a function of time. In addition, we observe that for large value of $DE_{0}$ we get correspondingly large value of average work.
\section{Summary and Conclusion}\label{section5}
\noindent A system of water molecules sparcely placed on a lattice in contact with a heat bath of temperature $T$ is immersed in a strong electric field $E_{0}$. A weak AC electric field is additionally switched on while recording its energy value at the initial equilibrium state. After AC field acted for a span of time $\tau$, then is switched off and its corresponding energy value recorded. Taking a complete cycle operation, the protocol is repeated for a large enough amount of time. This repeated measurement gave us the distribution of work by the external agent on the system. With this finite-time process, we managed to get the probability of work distribution and evaluated its different properties such as the mean work and average work as a function of time. \\
In conclusion, such sparcely confined water molecules are realizable in an experiment as done few years back \cite{dressel2018quantum} and would be interesting to carry out an experiment in line with our suggest to check our results.
\\
\noindent \textit{Acknowledgments:  Y.B. would like to thank  Wolkite University for financial support during his work.} 
\section*{Author contributions}
YB conception and design of study. YB performed the analytic  calculations and numerical results. YB, and  MB analysing and interpreting the results. YB, MB, and MM drafted manuscript preparation. All authors reviews the results and approved the final version of the manuscript. \\
\textbf{Funding}: The authors declare that they have no known competing financial interests.\\
\textbf{Conflict of Interest}: The authors declare no conflict of interest.\\

\section*{Data Availability Statement}
This manuscript has no associated data or the data will not be deposited.

\bibliographystyle{spphys}
\bibliography{mybibfile.bib}

\begin{thebibliography}{10}
\providecommand{\url}[1]{{#1}}
\providecommand{\urlprefix}{URL }
\expandafter\ifx\csname urlstyle\endcsname\relax
  \providecommand{\doi}[1]{DOI \discretionary{}{}{}#1}\else
  \providecommand{\doi}{DOI \discretionary{}{}{}\begingroup
  \urlstyle{rm}\Url}\fi

\bibitem{callen1985thermodynamics}
H.~Callen, New York  (1985)

\bibitem{gibbs1902elementary}
J.W. Gibbs, \emph{Elementary principles in statistical mechanics: developed
  with especial reference to the rational foundations of thermodynamics} (C.
  Scribner's sons, 1902)

\bibitem{feynman1998statistical}
R.~Feynman, \emph{Statistical Mechanics: A Set Of Lectures}.
\newblock Advanced Books Classics (Avalon Publishing, 1998).
\newblock \urlprefix\url{https://books.google.com.et/books?id=Ou4ltPYiXPgC}

\bibitem{zhong1996effect}
W.~Zhong, D.~Vanderbilt, Physical Review B \textbf{\textbf{53}}(9), 5047 (1996)

\bibitem{liphardt2002equilibrium}
J.~Liphardt, S.~Dumont, S.B. Smith, I.~Tinoco, C.~Bustamante, Science
  \textbf{\textbf{296}}(5574), 1832 (2002)

\bibitem{rousseau1999role}
R.~Rousseau, D.~Marx, The Journal of chemical physics
  \textbf{\textbf{111}}(11), 5091 (1999)

\bibitem{schollwock2011density}
U.~Schollw{\"o}ck, Annals of physics \textbf{\textbf{326}}(1), 96 (2011)

\bibitem{miller2017time}
H.J. Miller, J.~Anders, New Journal of Physics \textbf{\textbf{19}}(6), 062001
  (2017)

\bibitem{keyl2002fundamentals}
M.~Keyl, Physics reports \textbf{\textbf{369}}(5), 431 (2002)

\bibitem{elouard:hal-01170581}
C.~Elouard, A.~Auff{\`e}ves, M.~Clusel, {Stochastic thermodynamics in the
  quantum regime} (2015).
\newblock \urlprefix\url{https://hal.archives-ouvertes.fr/hal-01170581}.
\newblock Working paper or preprint

\bibitem{chou2011non}
T.~Chou, K.~Mallick, R.~Zia, Reports on progress in physics
  \textbf{\textbf{74}}(11), 116601 (2011)

\bibitem{crooks2000path}
G.E. Crooks, Physical review E \textbf{\textbf{61}}(3), 2361 (2000)

\bibitem{jarzynski1997nonequilibrium}
C.~Jarzynski, Physical Review Letters \textbf{\textbf{78}}(14), 2690 (1997)

\bibitem{jarzynski1997equilibrium}
C.~Jarzynski, Physical Review E \textbf{\textbf{56}}(5), 5018 (1997)

\bibitem{crooks1998nonequilibrium}
G.E. Crooks, Journal of Statistical Physics \textbf{\textbf{90}}(5-6), 1481
  (1998)

\bibitem{talkner2007fluctuation}
P.~Talkner, E.~Lutz, P.~H{\"a}nggi, Physical Review E \textbf{\textbf{75}}(5),
  050102 (2007)

\bibitem{crooks1999entropy}
G.E. Crooks, Physical Review E \textbf{\textbf{60}}(3), 2721 (1999)

\bibitem{ribeiro2016quantum}
W.L. Ribeiro, G.T. Landi, F.L. Semi{\~a}o, American Journal of Physics
  \textbf{\textbf{84}}(12), 948 (2016)

\bibitem{crooks2008jarzynski}
G.E. Crooks, Journal of Statistical Mechanics: Theory and Experiment
  \textbf{\textbf{2008}}(10), P10023 (2008)

\bibitem{shen2016quantum}
S.P. Shen, J.C. Wu, J.D. Song, X.F. Sun, Y.F. Yang, Y.S. Chai, D.S. Shang, S.G.
  Wang, J.F. Scott, Y.~Sun, Nature communications \textbf{\textbf{7}}(1), 1
  (2016)

\bibitem{rowley2014ferroelectric}
S.~Rowley, L.~Spalek, R.~Smith, M.~Dean, M.~Itoh, J.~Scott, G.~Lonzarich,
  S.~Saxena, Nature Physics \textbf{\textbf{10}}(5), 367 (2014)

\bibitem{dressel2018quantum}
M.~Dressel, E.S. Zhukova, V.G. Thomas, B.P. Gorshunov, Journal of Infrared,
  Millimeter, and Terahertz Waves \textbf{\textbf{39}}(9), 799 (2018)

\bibitem{sutmann1998structure}
G.~Sutmann, Journal of Electroanalytical Chemistry \textbf{\textbf{450}}(2),
  289 (1998)

\end{thebibliography}

\end{document}